\input phyzzx.tex
\tolerance=1000
\voffset=-0.0cm
\hoffset=0.7cm
\sequentialequations
\def\rl{\rightline}

\def\t1{{\tilde 1}}

\REF{\DIM}{N. Arkani-Hamed, S. Dimopoulos and G. Dvali,  Phys. Rev. Lett.
{\bf B429} (1998) 263, hep-ph/9803315;
I.~Antoniadis, N.~Arkani-Hamed, S.~Dimopoulos and G.~Dvali,
Phys. Lett. {\bf B436}, 257 (1998), hep-ph/9804398.}
\REF{\OLD}{I. Antoniadis, Phys. Lett. {\bf B246} (1990) 377; J. Lykken,
 Phys. Rev. {\bf D54} (1996) 3693, hep-th/9603133; I. Antoniadis and K.
Benakli, Phys. Lett. {\bf B326} (1994) 69;
I. Antoniadis, K. Benakli and M. Quiros, Nucl. Phys. {\bf B331} (1994) 313.}
\REF{\LYT}{ D.H. Lyth,  Phys. Lett. {\bf B448} (1999) 191, hep-ph/9810320.}
\REF{\KL}{ N. Kaloper and A. Linde, hep-th/9811141.}
\REF{\LIN}{A. Linde, Phys. Lett. {\bf B259} (1991)38; Phys. Rev. {\bf D49}
(1994) 748.}
\REF{\NIM}{N. Arkani-Hamed, S. Dimopoulos, N. Kaloper and J. March-Russell,
hep-ph/9903239.}
\REF{\BEN}{K. Benakli, hep-ph/9809582; K. Benakli and S. Davidson,
hep-ph/9810280.}
\REF{\MR}{ M. Maggiore and A. Riotto, hep-th/9811089.}
\REF{\DIE}{K.R. Dienes, E. Dudas, T. Ghergetta, A. Riotto, hep-ph/9809406.}
\REF{\DT}{G. Dvali and S.-H. Tye, hep-ph/9812483.}
\REF{\CGT}{C. Csaki, M. Graesser and J. Terning, hep-ph/9903319.}
\REF{\CLI}{J. Cline, hep-ph/9904495.}
\REF{\RIO}{A. Riotto, hep-ph/9904485.}
\REF{\DVA}{G. Dvali, hep-ph/9905204.}
\REF{\AFIV}{G. Aldazabal, A. Font, L. Ibanez and G. Violero, hep-th/9804026.}
\REF{\IMR}{L. Ibanez,  C. Munoz and S. Rigolin, hep-ph/9812397.}
\REF{\HBD}{E. Halyo, {\it Phys. Lett.} {\bf B387}  (1996) 43, hep-ph/9606423;
P. Binetruy and G. Dvali, {\it Phys. Lett.} {\bf B388}  (1996) 241,
hep-ph/9606342.}
\REF{\CUL}{S. Cullen and M. Perelstein, hep-ph/9903422.}
\REF{\EDI}{E. Halyo, hep-ph/9901302.}
\REF{\EXP}{N. Arkani-Hamed, S. Dimopoulos and G. Dvali,  hep-ph/9807344.}
\REF{\HAL}{E. Halyo, hep-ph/9904432.}

\singlespace
\rl{SU-ITP-99-21}
\rl{hep-ph/9905244}
\rl{\today}
\pagenumber=0
\normalspace
\medskip
\bigskip
\titlestyle{\bf{ D--term Inflation at the TeV Scale and Large Internal
Dimensions}}
\smallskip
\author{ Edi Halyo{\footnote*{e--mail address: halyo@dormouse.stanford.edu}}}
\smallskip
\centerline {Department of Physics}
\centerline{Stanford University}
\centerline {Stanford, CA 94305}
\smallskip
\vskip 2 cm
\titlestyle{\bf ABSTRACT}
We show that D--term inflation at the TeV scale is possible in the presence of
large internal dimensions. This requires very small couplings and masses which
arise from the large size of the compact dimensions. We show that acceptable
number of e-folds, magnitude and spectrum of density perturbations $\delta
\rho/\rho$ and $\eta$ can be obtained in this scenario at the price of a
reheating temperature
which is too high. Demanding an acceptable $T_R$ results in very small density
perturbations.

\singlespace
\vskip 0.5cm
\endpage
\normalspace

\centerline{\bf 1. Introduction}
\medskip

The existence of large (e.g. from micron to milimeter) internal dimensions is a
very exciting possibility that was raised recently[\DIM,\OLD]. This requires
that the higher dimensional gravitational scale be around the TeV scale. In
particular it means there must be string theories with $M_s \sim TeV$. In this
scenario, our world lives on a brane and gravity propagates only in the bulk.
The weakness of gravity with respect to the gauge interactions is a result of
the large internal dimensions. This scenario which is drastically different
than previously considered cases requires a complete revision of cosmology
above the nucleosynthesis scale. In particular, inflation which is so
successful in solving the horizon, flatness, monopole etc. problems of the
Big Bang cosmology must be realized in a novel way in this framework.

The possibility of inflation around the TeV scale in the presence of large
internal dimensions was first examined in [\LYT,\KL]. There it was shown that
chaotic inflation at the TeV scale is almost impossible whereas hybrid
inflation around the TeV
scale requires very small couplings and/or  masses. For example, hybrid
inflation with the
potential[\LIN]
$$V(\phi,\sigma)=g^2(M^2-\sigma^2)^2+m^2\phi^2 + \lambda^2 \phi^2 \sigma^2
\eqno(1)$$
gives the correct density perturbations for $g \sim \lambda \sim 1$ and $M \sim
1 ~TeV$
only if the mass of the inflaton is very small, i.e. $m^2 \sim 10^{-10}~eV$.
This mass is six orders of magnitude smaller than the Hubble constant $H \sim
10^{-4}~eV$ and is in general difficult
to obtain. A possible inflationary scenario was obtained in [\NIM] by
asymmetric inflation in the bulk and on the brane which occurs before the
internal dimensions are stabilized. Recent work
on related subjects appeared in [\BEN-\DVA].

In this letter, we study D--term inflation around the TeV scale in the presence
of large internal dimensions. We assume that the internal dimensions are
stabilized before inflation. In fact this assumption is crucial  because the
large dimensions result in very small couplings and masses
which make this scenario possible. We show that  we can obtain the acceptable
number of e-folds and magnitude and spectrum of density fluctuations quite
naturally. We comment on the prospects for obtaining an acceptable reheating
temperature.

\bigskip
\centerline{\bf 2. D-term Inflation at the TeV Scale}
\medskip

We now consider D--term inflation around the TeV scale in the presence of large
internal dimensions. For concreteness we study the case with only two large
dimensions but our
results can be easily generalized to the cases with more than two large
dimensions. Also since
models with large internal dimensions are naturally realized as type I string
orbifolds or type IIB orientifolds[\AFIV,\IMR], we consider D--term inflation
in these string models[\EDI].
D--term inflation
is a specific realization of hybrid inflation in which the vacuum energy during
inflation is dominated by a D--term rather than an F--term[\HBD].
For example, consider a type I string orbifold with an anomalous D--term
contribution to the potential
$$V_D=g^2|-|\sigma|^2+M^2|^2 \eqno(2)$$
and a tree level superpotential
$$W=\lambda \phi \sigma^2 \eqno(3)$$
Here $\sigma$ is the trigger field which carries $-1$ charge under the
anoumalous $U(1)_A$ symmetry whereas $\phi$ is the inflaton which is neutral.
The gauge coupling $g$ and the Yukawa coupling $\lambda$ are
not fixed; in particular they can be very small. As we will see there is a
natural reason for the smallness of these couplings. $M$ is a twisted modulus
field of the type I orbifold model and
its VEV gives the scale of the anomalous D--term. In particular this VEV can be
larger than the string scale which is about $50 ~TeV$[\CUL] for two large
dimensions. The superpotential gives rise to an F--term contribution in the
potential
$$V_F=\lambda^2 \phi^2 \sigma^2 +\lambda^2 \sigma^4 \eqno(3)$$
We see that $V=V_F+V_D$  contains two of the three terms required for hybrid
inflation.  Both of the above anomalous D--term and  the superpotential arise
generically in type I orbifold models.

The above potential $V=V_F+V_D$ has two minima; one at $\sigma=M$ and $\phi=0$
and the other at $\sigma=0$ and $\phi$ free. Inflation occurs for the initial
conditions of large $\phi$
and small $\sigma$. Then the vacuum energy is dominated by the D--term and is
given by
$$V_0\sim{H^2  M_P^2} \sim g^2 M^4 \eqno(4)$$
The masses for the two scalars are given by
$$m_{\sigma}^2=\lambda^2 \phi^2-2g^2 M^2 \eqno(5)$$
and
$$m_{\phi}^2=\lambda^2 \sigma^2 \eqno(6)$$
We see that $m_{\sigma}^2$ is positive for $\phi>{\sqrt 2}
gM/\lambda=\phi_{cr}$. For $\phi>\phi_{cr}$ we find
$m_{\sigma}^2>H$. As a result, $\sigma$ settles to its minimum at $\sigma=0$
very fast.
At the minimum the tree level inflaton mass vanishes, $m_{\phi}=0$.
However, the third term required for hybrid inflation which is the mass term
for $\phi$ arises from the one--loop contribution to the potential due to SUSY
breaking during inflation.
$$V_1=g^2M^4 \left(1+{g^2 \over {16\pi^2}} log{\lambda^2 \phi^2 \over
\mu^2}\right) \eqno(7)$$
Here $\mu$ is the renormalization scale which does not affect the physics.
$V_1$ leads to a mass for $\phi$ given by
$$m_{\phi}^2={g^4 M^4 \over {16 \pi^2 \phi^2}} \eqno(8)$$

Now that we have all the terms required for D--term inflation we need to fix
the parameters
of the model, e.g. $g,\lambda$ and $M$. The first requirement for inflation is
the slow--roll
condition
$$m_{\phi}^2 <H^2 \sim {g^2M^4 \over M_P^2} \eqno(9)$$
Since $M<<M_P$ in  this TeV scale scenario one needs an extremely small
inflaton mass for
inflation. From the form of the inflaton mass in eq. (8) we see that this is
possible only if the
gauge coupling is extremely small. This is not uncommon in models with very
large internal dimensions. For example consider a gauge theory in six
dimensions with a large internal torus.
Then the relation between the four and six dimensional  gauge couplings is
$$g_4^2={g_6^2 \over {R^2M_s^2}} \eqno(10)$$
The same holds for the Yukawa coupling $\lambda$. Thus we see that for a large
enough compact two torus the four dimensional gauge and Yukawa couplings will
be extremely small.

For example consider a model in which there are two D5 branes overlapping over
three dimensions. The type I string theory is compactified over four small
dimensions of size $M_s^{-1}$ and two large dimensions given by
$$R^2 \sim {M_P^2 \over M_s^4} \sim 2 \times 10^{17}~GeV^{-2} \eqno(11)$$
The three dimensions common to both branes are noncompact and describe the
world. We assume that one D5 brane is wrapped around a large two torus whereas
the other one is wrapped around a string size torus.
The Standard Model degrees of freedom come from the D5 brane wrapped over the
small two torus with gauge couplings of $O(1)$. Now if $\sigma$ lives on the D5
brane wrapped around the large dimensions and also the anomalous $U(1)_A$ comes
from this brane,
then the four dimensional gauge coupling, i.e. $g^2$ and will be suppressed by
the
factor $R^2 M_s^2 \sim 4 \times 10^{26}$. Assuming that the six dimensional
gauge coupling is $O(1)$ we get $g^2  \sim 3 \times 10^{-27}$. On the other
hand, if the inflaton $\phi$ lives on the D5 brane wrapped around the small
torus, then the Yukawa coupling will be of
$O(1)$, $\lambda \sim 1$, since there is no suppression factor due to the
volume of the two torus. In fact this is exactly what happens in orbifolds of
type I strings (or orientifolds of type IIB strings) for
gauge and Yukawa couplings.  
We see that the extreme smallness of the gauge coupling is a direct result of
the large internal dimensions whereas the large value of the Yukawa couling is
a result of the small torus.
We also need to know the value for $M$. The VEV of $M$ is a free parameter ($M$
is a twisted modulus[\HAL]) before SUSY breaking due to the nonzero vacuum
energy and it is perfectly reasonable for it to be larger than the string scale
as long as this does not lead to an energy density higher than $M_s^4$. 
For the moment we leave $M$ free but note that it will be much larger than $M_s
\sim 50~TeV$. Similarly $\phi$ can be much larger than $M_s$. We take
for the large initial value of  the inflaton $\phi \sim 6 \times 10^6~GeV$.
Later we will see that this is more than enough for 60 e--foldings.

Initially using eq. (8) we get  for the inflaton mass $m_{\phi} \sim 7 \times
10^{-5}~eV$ whereas the Hubble constant is
$H \sim 10^{-1}~eV$. The inflaton mass is four orders of magnitude smaller than
$H$ and therefore inflation happens in this scenario. Inflation continues as
long as $m_{\phi}<H$ and stops when
$m_{\phi} \sim H$. Then $\phi$ rolls to its minimum at $\phi=0$. On the other
hand,  $m_{\sigma}^2$ becomes negative and $\sigma$ rolls from the (now)
maximum at $\sigma=0$ to the new minimum at $\sigma=M$. The vacuum energy
becomes zero, inflation ends
and SUSY is restored. The number of e-folds during inflation is given by
$$N \sim \left(\phi_i \over M_P \right)^2 {4\pi^2 \over g^2} \eqno(12)$$
where $\phi_i$ is the initial value of the inflaton. We obtain $N \sim 4 \times
10^5$ which is certainly enough inflation to solve the horizon and flatness
problems.
Now for inflation we are interested in quantities corresponding to the latest
60 e--folds. From eq. (8) 
we find that the magnitude of the inflaton $\phi$ corresponding to $N \sim 60$
is $\phi \sim 7 \times 10^4~GeV$ slightly above the string scale $M_s$.
Note that this value is much smaller than $\phi_i$ we assumed above but any
$\phi_i > 7 \times 10^4~GeV$ will give acceptable inflation.
Substituting the value of $\phi$ from eq. (12) into eq. (8) we find that the
inflaton mass is
smaller than the Hubble constant by a factor of $\sqrt {4N} \sim 12$
independently of $g^2$ and $M$ at the time of the latest 
60 e--foldings. Later once we fix $M$ we will find the values for the inflaton
mass and Hubble constant at 60 e--folds.

The magnitude of density perturbations is given by
$${\delta \rho \over \rho} \sim {\lambda g^2 M^5 \over {M_P^3 m_{\phi}^2}}
\eqno(13)$$
Using the values for $g,\lambda$ $m_{\phi}$ from eq. (8) and $\phi$ from eq.
(12) we obtain $\delta \rho/ \rho \sim 4 \lambda N M/M_P$. Untill now $M$ was
not fixed and we see that
it has to be $M \sim 4 \times 10^{10}~GeV$ in order to get the correct density
perturbations
of $\delta \rho/ \rho \sim 10^{-5}$ at 60 e--foldings
obtained from the COBE data. We see that $M$ is many orders of magnitude larger
than $M_s$ but this is acceptable since the vacuum energy density $g^2 M^4 \sim
10^{-3} M_s^4$.
Note that the magnitude of the density perturbations is independent of the very
small gauge coupling constant
$g^2$ but depends on the Yukawa coupling $\lambda$.
Now that we have fixed $M$ we can calculate the $m_{\phi}^2$ and $H^2$ from
eqs. (8) and (9).
We find that at 60 e--foldings $m_{\phi} \sim 10^{-2}~eV$ whereas $H \sim
10^{-1}~eV$.

A second requirement for inflation is the condition on the spectrum of density
perturbations
$$\eta \sim \sqrt{g^2 \over {16 \pi^2}} {M_P \over \phi} <0.2 \eqno(14)$$
This has to be satisifed at the time of 60 e--foldings, i.e. for $\phi \sim 7
\times 10^4~GeV$. Substituting $g^2$ and $\phi$ we find $\eta \sim 0.06$.
Note that the extreme smallness of $g^2$ is crucial for this result.

We find that the D--term inflation with large internal dimensions satisfies all
the requirements of acceptable inflation. We stress that the smallness of the
gauge coupling plays a crucial role in this scenario. 
The flatness of the inflaton potential and satisfying the constraints on
$\delta \rho/\rho$ and $\eta$ is a direct consequence of the very small gauge
coupling and the large value of $M$ (compared to $M_s$). 
On the other hand,
extremely small gauge couplings are generic to models with large internal
dimensions. Therefore,
if inflation occurs after the internal dimensions are stabilized at large
values the D--term inflation scenario becomes very natural.

Finally we would like to find out the reheating temperature $T_R$. In [\EXP] it
was shown that if one does not want to heat up the bulk by graviton production
one needs $T_R \sim 10~MeV$.
At the end of inflation the inflaton mass becomes
$$m_{\phi}^2 \sim \lambda^2 \sigma^2 \sim \lambda^2 M^2 \sim 10^{21}~GeV^2
\eqno(15)$$
which gives $m_{\phi} \sim 4 \times 10^{10}~GeV$. The main decay channel for
the inflaton is to two fermionic $\sigma$'s due to the tree level
superpotential in eq. (3). (Note that after inflation the fermionic $\sigma$'s
are massless since $\phi=0$.)
The rate of this decay is given by $\Gamma \sim \lambda^2 m_{\phi} /25 \sim
10^9~GeV$ which is much larger than the Hubble constant. Therefore, after
inflation ends all the vacuum energy is converted into heat very efficiently.
As a result,
$$T_R^4 \sim g^2 M^4 \eqno(16)$$
This gives $T_R \sim \sqrt{g} M \sim 10^4 ~GeV$ which is six orders of
magnitude larger than what is needed. We see that the high  $T_R$ is a result
of the very large value for $M$ which was required for the correct amount of
density perturbations.
If we  take $M \sim 10^5~GeV$ then we get $T_R \sim 10^{-2}~GeV$
which is about the normalcy temperature but then the density perturbations are
too small, i.e. $\delta \rho/ \rho \sim10^{-10}$.

\bigskip
\centerline{\bf 3. Discussion}
\medskip

In this letter we considered D--term inflation around the TeV scale with large
internal dimensions. We showed that if the internal dimensions are stabilized
before
inflation they lead to very small couplings and masses which are crucial for
the success of the scenario. We find the fact that we obtained realistic values
for the magnitude and spectrum of density perturbations,
quite naturally very promising. It is highly interesting that the main
requirement of the scenario which is small couplings are in fact a consequence
of the large compact dimensions. Thus once the internal dimensions are
stabilized at their large values there is not a new naturalness problem for the
couplings. We also  found that an acceptable reheating temperature is hard to
obtain in this scenario.

The main drawback of this scenario is the fact that we need the large internal
dimensions to be stabilized before inflation. In fact in ref. [\KL] it was
shown that even in that case there is a naturalness problem. Inflation occurs
at times of order $H^{-1}$ whereas the natural time scale is $M_s^{-1}$ which
is many orders of magnitude smaller. The requirement for the homogeniety of the
universe between these two times is very hard to explain. Nevertheless, we
think that the succsess of D--term inflation at the TeV scale merits further
study of this problem.

For simplicity we only considered the case of two large dimensions in this
letter. However, all of our results can be generalized easily to cases with
more than two large dimensions. For example if there are three large dimensions
one can consider the anisotropic case where two large dimensions are of the
size given above while the third is very small. Assuming the same values for
$M$ and the coupling constants we trivially get the same results. Note that in
this case the lower bound on the string scale is much smaller but the value of
$M$ is not necessarily related to $M_s$.

\bigskip
\centerline{\bf Acknowledgements}

We thank Raphael Bousso and especially Nemanja Kaloper for very useful
discussions.

\refout
\end
\bye